# High-precision measurement of total fission cross sections in spallation reactions of $^{208}$Pb and $^{238}$U

K.-H. Schmidt[1], B. Jurado[2], R. Pleskač[1], M. V. Ricciardi[1], J. Benlliure[3], T. Enqvist[4], F. Farget[5], A. Bacquias[1,*], L. Giot[1,†], V. Henzl[1,‡], D. Henzlova[1,‡], A. Kelić-Heil[1,§], S. Leray[6], S. Lukić[1,**], Son Nguyen Ngoc[1], A. Boudard[6], E. Casarejos[3], M. Fernandez[3], T. Kurtukian[2], P. Nadtochy[1,††], D. Perez[3], C. Schmitt[5]

[1] *GSI-Helmholtzzentrum für Schwerionenforschung GmbH, D-64291 Darmstadt, Germany*
[2] *CENBG, CNRS/IN2 P3, Chemin du Solarium B.P. 120, F-33175 Gradignan, France*
[3] *Universidad de Santiago de Compostela, E-15706, Spain*
[4] *Oulu Southern Institute and Department of Physics, University of Oulu, Finland*
[5] *Grand Accélérateur National d'Ions Lourds, BP 55027, F-14076 Caen Cedex 05, France*
[6] *DSM/IRFU/CEA, 91191 Gif-sur-Ivette, France*

**Abstract:** Total cross sections for proton- and deuteron-induced-fission of $^{208}$Pb and $^{238}$U have been determined in the energy range between 500 MeV and 1 GeV. The experiment has been performed in inverse kinematics at GSI Darmstadt, facilitating the counting of the projectiles and the identification of the reaction products. High precision between 5% and 7% has been achieved by individually counting the beam particles and by registering both fission fragments in coincidence with high efficiency and full Z resolution. Fission was clearly distinguished from other reaction channels. The results were found to deviate by up to 30 % from Prokofiev's systematics on total fission cross sections. There is good agreement with an elaborate experiment performed in direct kinematics.

## 1. Introduction

Since the accelerator-driven system (ADS) is considered as an option for the incineration of radioactive waste [1,2,3], intense effort has been made in providing experimental data on interactions of intermediate-energy ($E \approx 100$ to 1000 MeV) protons and neutrons with the neutron-production target, with construction material, and with materials that undergo transmutation in the ADS. Because of the variety of target nuclei and the wide range of energy of the beam particles, as well as the large number of open reaction channels, it is impossible to measure all needed data. Thus, theoretical models and nuclear-reaction codes, based on the models, are needed. In 1997, the systematic intercomparison [4] of codes for calculation of radioisotope production observed in irradiation of different target material by intermediate-energy protons showed that the predictive power of the codes available at that time was not better than ±50%, while there were cases of disagreements of orders of magnitude between the calculated and measured data. In the mean time, the situation has improved considerably [5,6], mainly due to the experimental and theoretical work, started by a European initiative and performed in the frame of the HINDAS [7] and the n_TOF [8] projects and later in NUDATRA [9]. In particular, the experimental knowledge on the production of individual nuclides in charged-particle induced spallation reactions has improved substantially thanks to the results of an experimental campaign executed in inverse kinematics at GSI Darmstadt [10] with a high-resolution magnetic spectrometer. Also improved codes [11, 12] have been developed on the basis of this new generation of experimental results.

---

[*] Present address: CNRS-IPHC Strasbourg, France
[†] Present address : Subatech - Ecole des Mines de Nantes
[‡] Present address: Los Alamos National Laboratory, NM, USA
[§] Contact author: a.kelic@gsi.de
[**] Present address: Vinča Institute of Nuclear Sciences
[††] Present address: Omsk State University, Department of Theoretical Physics, RU-644077 Omsk, Russia

The situation has not so much improved, however, concerning total fission cross sections. Experimental uncertainties are often rather large, and the results of different experiments severely contradict each other in many cases, as being documented in a systematic overview by Prokofiev [13]. Previous experiments performed at GSI using the fragment separator FRS (see below) did not reach the high precision for spallation-fission products which they reached for spallation-evaporation residues, due to the low transmission of the used spectrometer for the large-emittance fission-fragment beams. Fission reactions may have a significant effect on the performance of the spallation target of the ADS, in particular on the production of radioactive and/or chemically hazardous materials in the target. Precise knowledge on total fission cross sections is even more important since both (p,f) and (n,f) cross-sections are used as standards for flux measurements in the energy region above 20 MeV, which are important for ADS applications.

In the present work, we report on the first results of a new generation of high-precision measurements of total fission cross sections in spallation reactions of $^{208}$Pb and $^{238}$U at energies between 500 and 1000 $A$ MeV. The experiments were performed with a novel experimental approach in inverse kinematics using a full-acceptance detection system. This technique has decisive advantages and copes with several problems of most conventional direct-kinematics experiments performed up to now.

## 2. Experimental methods
### 2.1 Experiments in direct kinematics
Conventional experiments on total fission cross sections are performed in direct kinematics. Heavy target materials are bombarded with protons or neutrons of the energy of interest, and the fission products are registered with appropriate detectors. A variety of experimental techniques have been applied to detect the fission fragments and obtain total fission cross sections, see [14] for a detailed discussion:

*The radiochemical method* was among the first ones to be applied in fission detection. Nowadays, high-resolution gamma detectors are used to identify the radioactive species after irradiation. The complex time evolution of the activity due to radioactive decay and the lack of radiation from stable isotopes lead to incomplete coverage of the full nuclide production and, thus, to large uncertainties. *Nuclear photoemulsions* also belong to the first generation of detection techniques. Restricted choice of target material, background problems, and tedious track counting lead to limited application in cross-section measurements. *Solid-state nuclear-track detectors* eventually mounted on both sides of a thin target allow detecting correlated fragments. Again, tedious track counting is required. *Ionization chambers* characterize fission quite well, because they measure energy and emission angle of both fragments. Due to their sensitivity to protons, the application is mostly limited to (n,f) reactions. *The parallel-plate avalanche counter* (PPAC) can be advantageous due to its excellent timing. *Semiconductor detectors* provide fragment energy combined with a good timing. However, the sensitivity to radiation damages limits their application. *Thin-film breakdown counters*, metal-oxide silicon capacitors, are quite promising due to their threshold properties and good timing.

Even with the variety of detection techniques available, the optimum features: Large solid angle, good timing, coincident detection of both fragments and information on the multiplicity, low background, high resolution in energy and/or mass, are difficult to reach simultaneously. In addition, it is problematic that the beam dose is measured independently from the detection of the fission fragments. Also, the short range of the fission fragments in the target material and their emission in the full solid angle make it difficult to achieve high-precision results and to avoid systematic uncertainties.

The present experimental knowledge on total fission cross sections is well documented by Prokofiev [13]. Although there are many data published with relatively small uncertainties, some of those are in severe contradiction. For many targets and in wide energy ranges, the typical uncertainties are still rather large, up to a factor of two or more.

*2.2 Innovative experimental set-up in inverse kinematics*

Acquiring high precision in these experiments relies on an accurate determination of the beam dose and an unambiguous identification and counting of fission products. Also the discrimination of residues formed in other kind of nuclear reactions and the precise knowledge of the efficiency of the fission detectors are critical issues. The use of inverse kinematics offers advantages in reaching these goals.

The inverse-kinematics method is based on the bombardment of a hydrogen or other light target with relativistic heavy projectiles. The reaction products are identified in-flight. Their multiplicities, emission angles and velocities provide information on the reaction kinematics.

The fission set-up developed at the GSI experimental facility and situated behind the fragment separator (FRS) [15] is shown in Figure 1. It consists of scintillation detectors [16], characterized by fast time response and high time resolution, and ionization chambers [17], which provide high-resolution information on the energy loss and give information on the trajectory of the ions. The first scintillator is used as the trigger of the data acquisition and as the start detector for the time-of-flight to the TOF wall. The target is surrounded by two MUltiple Sampling Ionization Chambers (MUSICs), which altogether serve as an active target. This active target is able to select nuclear reactions occurring in the target, and to eliminate those occurring in air or originating from any other layer of matter situated upstream the target. The double ionisation chamber (TWIN) and the TOF wall measure the nuclear charges of the fission fragments and their velocity vectors in space. Additionally, the two ionisation chambers situated before (MUSIC1) and after the target (MUSIC2) can give information on the horizontal ($x$) position and angle of the passing ions, while the TWIN chamber can give the information on the horizontal ($x$) and vertical ($y$) position and angle of the reaction products. As it will be detailed below, the set-up provides a detection efficiency for fission products of more than 90%. The losses are due to the vertical emittance of the primary beam and an eventual shift of the TWIN cathode with respect to the mean vertical position of the beam, which would give a small probability that both fission fragments pass through the same half of the TWIN chamber. Additionally, fission fragments moving very close to the cathode could have less active volume to ionise, causing some additional losses.

For the target, two options can be considered: (a) a liquid-hydrogen (respectively deuterium) target, or (b) a plastic target. The cross section in hydrogen or deuterium must be deduced from a differential measurement. In the case (a) an additional background measurement is performed with an empty target container, and in the case (b) two additional measurements with a carbon target and no target are necessary. The use of the liquid-hydrogen respectively deuterium target gives certainly a more reliable result both on the total fission cross section and on the nuclear charge distribution of the fission fragments.

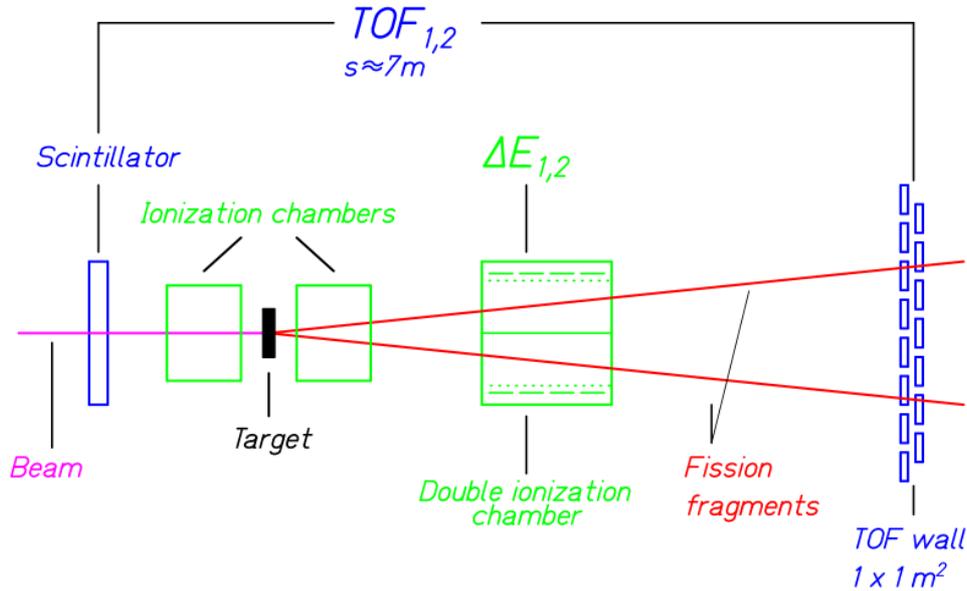

**Figure 1**: **Side view of the fission set-up mounted behind the FRS at GSI Darmstadt. Two ionization chambers (MUSIC1 and MUSIC2) with the target mounted in-between form an active target. A double ionization chamber (TWIN) and a TOF wall detect the two fission fragments.**

The present experimental approach has several essential advantages: The most important feature is that, due to the inverse kinematics, all fission fragments leave the target with high energy in a narrow cone in forward direction. In addition, the individual projectiles are registered and identified with the same detectors which also register the fission fragments. The angular range where fission fragments are emitted is fully covered. Further on, both fission fragments are registered and identified simultaneously, and their velocity vectors are determined. Finally, the multiplicity of the reaction products is accessible. All these features are crucial for obtaining total fission cross section with high precision.

## 3. Data analysis

### 3.1. Selection of the primary beam

For evaluating the fission cross section, the first task of the analysis consisted of determining the number of projectiles impinging on the target. In the present experiment, the projectiles were registered one by one by the ionization chamber MUSIC1 placed in front of the target. This was needed, since the projectiles delivered by the SIS18 accelerator might undergo a nuclear reaction in some layer of matter or in air before reaching the reaction target. In addition, it has to be assured that the projectiles hit the active area of the target. In each event, the ionization chamber MUSIC1 has recorded the energy loss of the passing ions. Events in which the primary beam has undergone nuclear reactions leading to a loss of protons before reaching the target could be discarded according to the energy-loss signal. Some reaction channels where only neutrons are removed cannot be distinguished from the non-interacting primary beam. Nevertheless, as these residues differ only by a few neutrons from the projectile, they have rather similar fissilities as the projectile, and their small contribution (less than 1 %) does not noticeably influence the result of the experiment. Figure 2 shows, in case of the reaction $^{208}$Pb (500 $A$ MeV) + $^{1}$H, the energy-loss spectrum recorded in the MUSIC1 chamber. Only the ions, which produce an energy-loss signal inside the indicated window, denoted by the two vertical lines, are considered. The two peaks directly below the selected one consist of projectile fragments with $Z$=81 and 80, respectively, and, to a great part, of incompletely stripped projectiles. The contribution of ions with $Z \neq 82$ to the events selected by the condition is estimated to be less than 0.3 %. Most of the lower energy-loss signals are produced by lighter elements, coming from nuclear reactions in some layer before the target. The shoulder above the selected peak is due to nuclear-charge-pickup reactions occurring before the projectiles reach the target. In most part of chapters 2 and 3, detailed information is given for the experiment $^{208}$Pb (500 $A$ MeV) + $^{1}$H as an example. The analysis procedure applied was very similar in the other cases; important deviations will be mentioned explicitly.

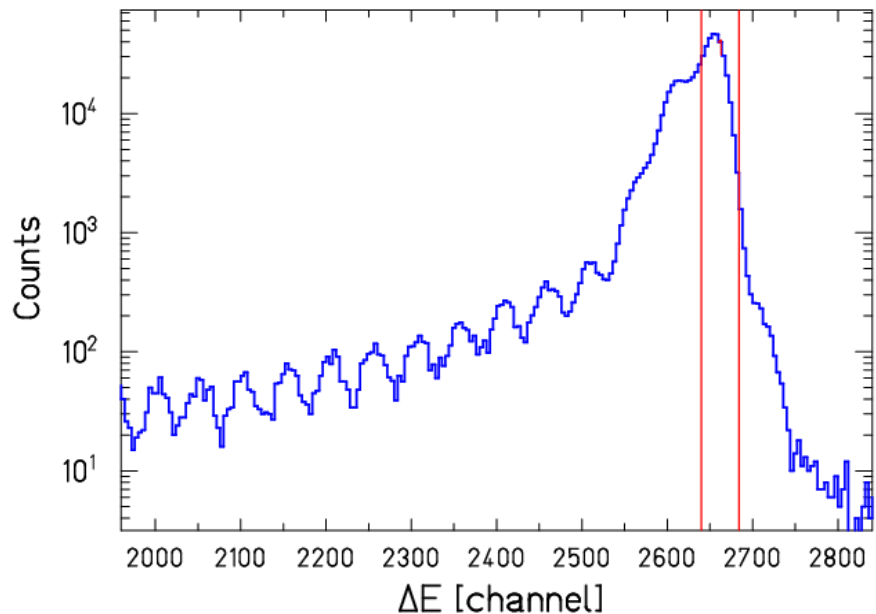

**Figure 2: Energy-loss spectrum of the ions recorded with the ionization chamber MUSIC1 placed directly in front of the target. Only the ions, which produce an energy-loss signal between the two red vertical lines, are counted as projectiles. The spectrum refers to the reaction $^{208}$Pb (500 $A$ MeV) + $^{1}$H.**

Since the diameter of the liquid target used in the present experiment was only 30 mm, an additional scintillation detector with a hole of 15 mm diameter was mounted in front of the target in order to detect projectiles which do not hit the target. Only those projectiles were considered, which did not produce any signal in the veto detector.

Reactions in the air section between the target and the TWIN ionization chamber were recognized by registering the energy-loss signal in the second ionization chamber MUSIC2 located directly behind the target. Reactions induced by the projectile particles in the target are situated inside the polygon lines drawn in Figure 3 on the two-dimensional energy-loss spectrum, while ions which reacted behind the target still delivered the full energy-loss signal in the second ionization chamber. In the following, only events situated inside the condition shown in Fig. 3 are considered.

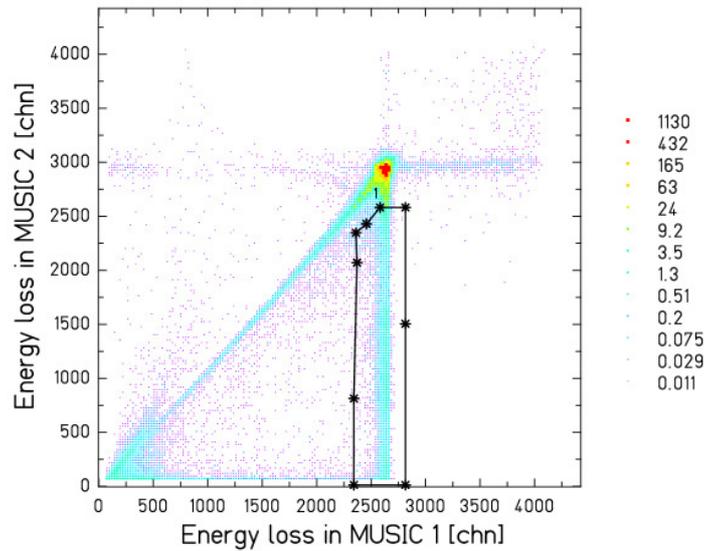

**Figure 3:** Two-dimensional representation of the energy-loss signals in the ionization chambers MUSIC1 directly in front of the target (horizontal axis) and MUSIC2 directly behind the target (vertical axis). The polygon line selects reaction residues produced in the target. The spectrum refers to the reaction $^{208}$Pb (500 $A$ MeV) + $^{1}$H.

*3.2. Identification of fission products*

The TWIN ionization chamber and the scintillator wall have the task to identify the fission products and provide information on their nuclear charge and angle. The TWIN ionization chamber registers the two fission products in most of the cases in the two separate gas volumes above and below the common cathode. A two-dimensional spectrum of the signals registered in the two parts of the TWIN chamber is shown in Figure 4. This spectrum is collected under the condition defined by the polygon window in Figure 3, i.e. after selecting events originating from reactions of the projectiles inside the target. Events populating the central peak in Figure 4 and selected by the polygon window correspond to fission events, while the fragments produced in fragmentation reactions and central collisions populate the edges of the spectrum. It is clear from the figure that the region populated by fission fragments is well separated from other reactions. Please note that the detection probability of the set-up does not fully reach 100% and thus not all fission events are contained inside the polygon window of Figure 4; as it will be shown later, this can be easily corrected for.

The pulse heights of the 16 separate anodes of the TWIN ionization chamber have been calibrated by connecting specially designed stripes at the anode plates to a pulse generator in a way that all calibrated pulse heights correspond to the same initial number of charge carriers. Considering that the ionization signal is proportional to the square of the nuclear charge of the fission fragments, the two-dimensional energy-loss spectrum (Figure 4) was converted into a $Z$ spectrum, see Figure 5. The absolute calibrations of the two sections of the TWIN ionization chamber used for Figure 4 was performed by determining the common scaling factor using the sum spectrum $Z_1 + Z_2$ of nuclear charges of the two fission fragments, see Figure 6. We assumed that the peak with highest $Z_1+Z_2$ value but with still relatively large intensity corresponds to the projectile atomic number $Z_p$, which in the present case amounts to 82, while the very weak peak to its right corresponds to fission after charge pickup. The uncertainty of this calibration is at most of one unit. The calibrated two-dimensional spectrum of the nuclear charges of the two fission fragments (Figure 5) shows a clear accumulation in the Z range expected for fission fragments produced in symmetric fission.

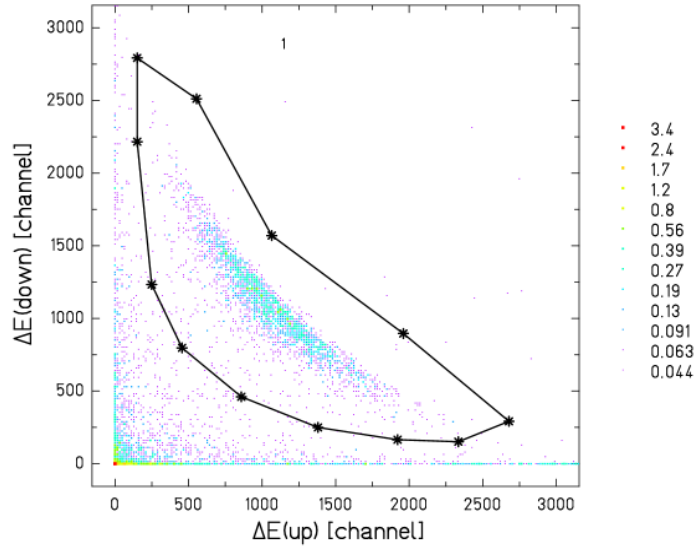

**Figure 4:** Two-dimensional spectrum of the energy-loss signals of nuclear-reaction products recorded by the TWIN ionization chamber. Signals of fission fragments lie inside the polygon window. The spectrum refers to the reaction $^{208}$Pb (500 $A$ MeV) + $^{1}$H.

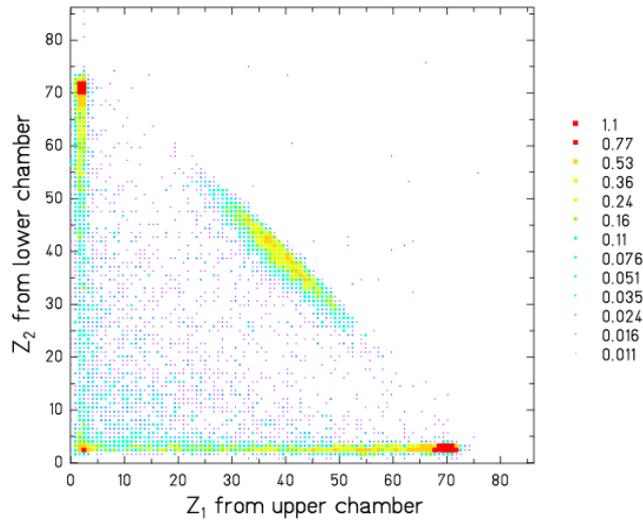

**Figure 5:** Two-dimensional spectrum of the nuclear charges of the two fission fragments recorded simultaneously by the TWIN ionization chamber. The spectrum refers to the reaction $^{208}$Pb (500 $A$ MeV) + $^{1}$H.

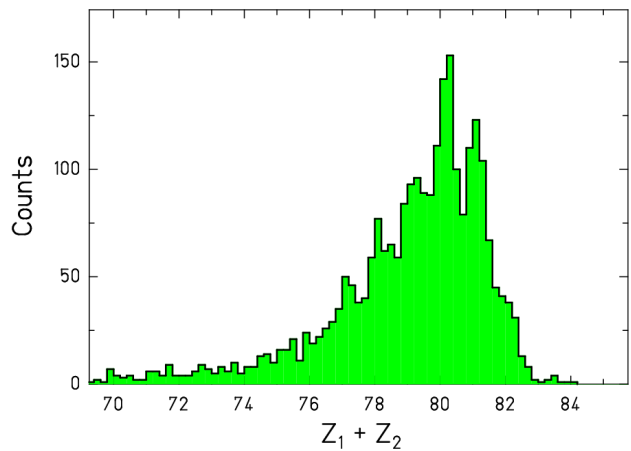

**Figure 6: Sum spectrum $Z_1 + Z_2$ of the nuclear charges of the two fission fragments. The different integer values of the charge sum are clearly visible. The spectrum refers to the reaction $^{208}$Pb (500 $A$ MeV) + $^1$H.**

### *3.3. Detection efficiency of fission events*

Due to the finite width of the primary beam in position and angle, fission products emitted close to the cathode plane may pass through the same part of the TWIN chamber. These events are not recognized as fission fragments but rather as fragmentation residues and are the reason that the detection probability of the set-up does not fully reach 100%. In order to be able to evaluate the losses, we first determined the angles of the trajectories of the ions. The TWIN chamber gives an absolute horizontal position, because the anodes have triangular shape and the calibration is directly given by the geometry of the anodes [18]. The vertical position can be inferred from the measured drift time, but a calibration is needed, since the drift velocity depends on several parameters like the purity of the gas and the temperature. This calibration can easily be deduced from a correlation with the paddle of the scintillator wall which is finally hit by the ion. Since the neighbouring paddles of the scintillator wall, having a width of 10 cm, overlap with 1/3 of their width, conditions on coincidences and anti-coincidences of the signals from different paddles characterize the vertical position of the ion in cells of 1/3 of paddle width. Figure 7 shows a spectrum of angles, deduced from drift-time differences, registered in the first and the last anodes of the two parts of the TWIN chamber in coincidence with different cells of the scintillator wall. Spectra corresponding to neighbouring cells are drawn in different colours. One can easily deduce the borders of the different cells, projected on the drift times in the TWIN chamber. In fact, this spectrum has been accumulated after applying the calibration, which is described in the following. In addition, the spectrum has been accumulated under the condition that exactly two cells of the scintillator wall were hit. This way, the borders of the different cells appear clearer in the spectrum. However, this induced a slight loss of fission fragments, which are undoubtedly identified in the TWIN correlation plot of figure 4. Already the width of the dip in the centre of the spectrum allows estimating the losses in the vicinity of the cathode to less than 10%.

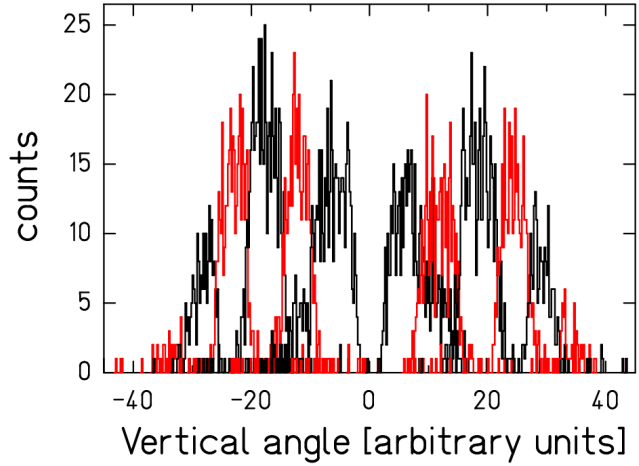

**Figure 7:** Vertical angles as deduced from drift-time difference of the last and the first anode of the TWIN detector, divided in sections according to the coincidences with different cells of the scintillator wall. Events corresponding to neighbouring cells are drawn in different colours. Negative angles were measured in the lower part, positive angles were measured in the upper part of the TWIN detector. The dip in the centre is caused by the efficiency losses in the vicinity of the cathode. The figure refers to the reaction $^{208}$Pb (500 $A$ MeV) + $^{1}$H.

The correlation of the borders of the cells of the scintillator wall and their projections on the drift-time differences are shown in Figure 8. The raw drift-time differences of the upper and the lower part of the TWIN detector were shifted by an additional constant each in order to obtain a consistent calibration.

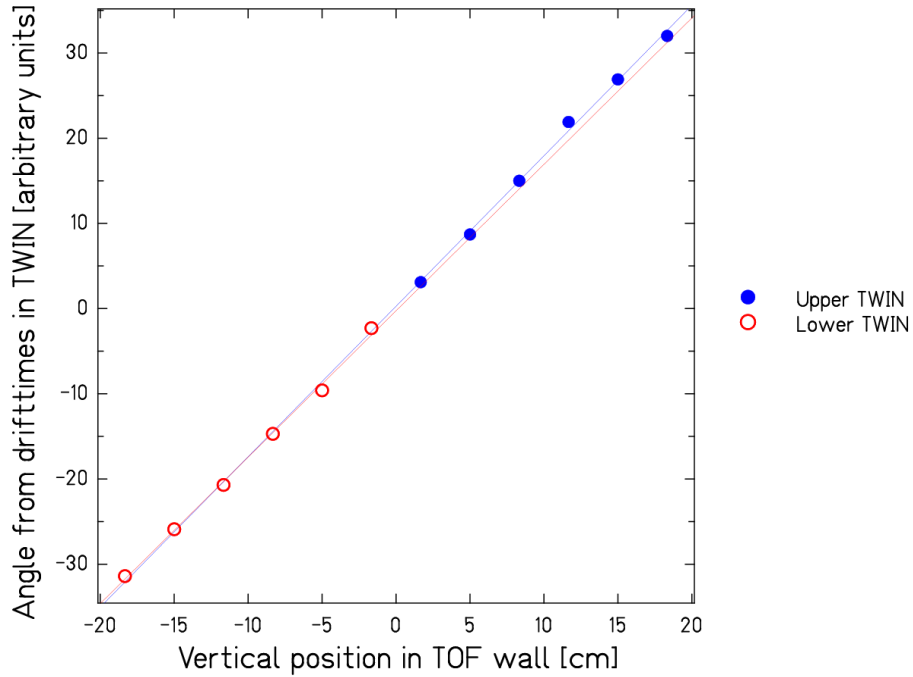

**Figure 8:** Vertical positions of the borders of the cells of the scintillator wall (horizontal axis) and their projections on the drift-time differences in the TWIN detector (vertical axis) after calibration. The drift-time differences in the two sections of the TWIN detector were shifted by a constant value each in order to obtain a consistent calibration.

After this calibration, a two-dimensional distribution of the angles perpendicular to the beam direction was accumulated for fission fragments. Figure 9 shows that this spectrum has the shape of a diffuse ring, as expected. The losses in the vicinity of the cathode of the TWIN detector are responsible for a slight reduction of counts for small vertical angles. The angles, indicated in arbitrary units, correspond approximately to mrad. There is a slight shift between the deduced horizontal positions in the two parts of the TWIN chamber, which might be a hint that the electrical field inside the chamber is not exactly vertical. This, of course, does not have any influence on the cross-section determination.

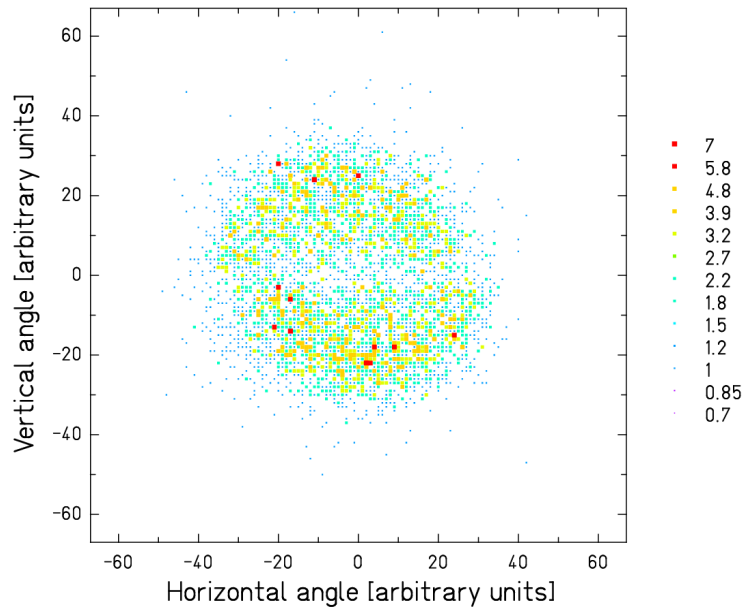

**Figure 9: Distribution of the angles in the plane perpendicular to the beam direction, gated on fission events. The figure refers to the reaction $^{208}$Pb (500 $A$ MeV) + $^{1}$H.**

An effort was made to determine accurately the detection efficiency of the TWIN chamber for fission fragments in the present experiment. For this purpose, the calibrated vertical angles of fission fragments were accumulated in a spectrum, see Figure 10. Figure 10 is equivalent to a projection of Figure 9 on the vertical axis. The fission fragments were selected by the two-dimensional energy-loss spectrum of Figure 4. If one neglects the fluctuations of the velocities of the fission fragments in the frame of the emitting source, the spectrum of Figure 10 should have the shape of a rectangular function. Obviously, the shape of the measured spectrum is similar to the one obtained by convoluting this rectangular function with a Gaussian distribution caused by the fluctuations in the velocities of the fission fragments, which are mostly due to the different mass splits. The fit with such a function is able to reproduce the measured spectrum rather well, except two features: The dip in the central region and a small asymmetry. As mentioned above, the dip results from the efficiency loss of the TWIN detector close to the cathode that we want to determine. The asymmetry of the measured spectrum has another origin. The separate spectra of the angles recorded in the upper and in the lower part of the TWIN detector reveal a malfunction of the upper part: The signals of some of the ions passing through the upper half of the TWIN detector very close to the cathode are not properly recorded by the corresponding time-to-amplitude converter, which results in wrong, partly negative angles‡‡. Using this information, we state that the correct spectrum is symmetric, being slightly lower at negative angles and slightly higher than the recorded one (full-line histogram in Figure 10) at positive angles. Thus the true spectrum is rather well represented by the fit curve. Having traced back the asymmetry of the spectrum to a wrong tracking information of some trajectories in the TWIN detector, the detection efficiency can safely be estimated by the magnitude of the dip near the centre to (94 ± 3) %.

---

‡‡ Note that the number of events with misidentified angles was about 1 % of all events, giving a small contribution of only 1.5% to the systematic uncertainty of the measured total fission cross section. This correction was not needed in case of $^{238}$U beam.

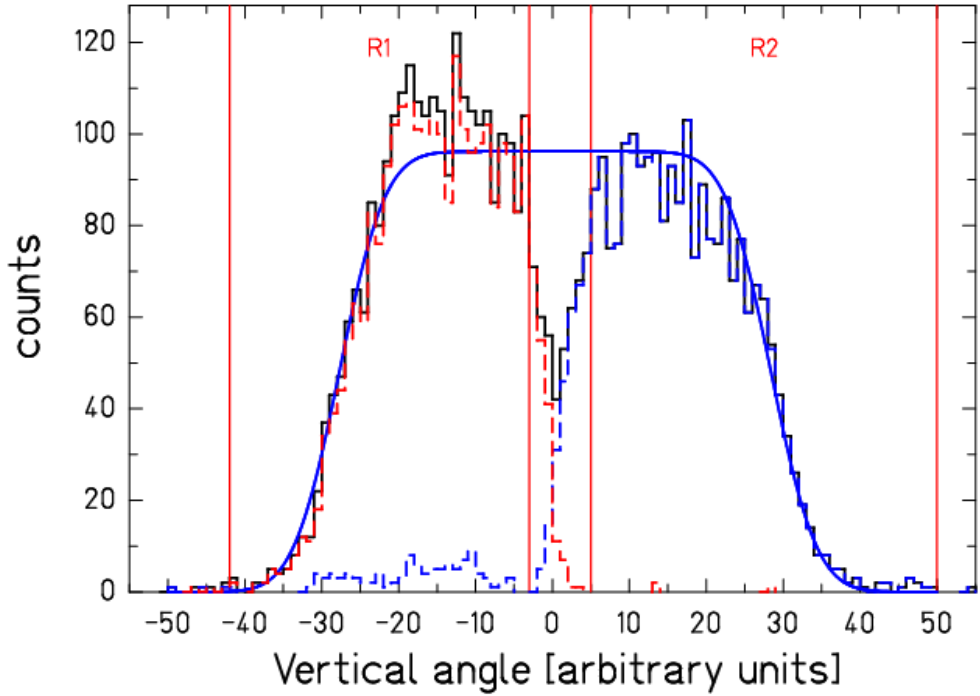

**Figure 10:** Spectrum of vertical angles of fission fragments (black full-line histogram). The angles registered in the lower (red dashed histogram) and in the upper half (blue dashed histogram) of the TWIN chamber are shown separately. The full line shows the result of a fit with a theoretical function, defined as the convolution of a rectangular function with a Gaussian (the full red vertical lines define windows used in the fit procedure). The spectrum refers to the reaction $^{208}$Pb (500 $A$ MeV) + $^{1}$H. For more details, see text.

As mentioned above, all results shown in Figures 2 to 10 refer to the reaction $^{208}$Pb (500 $A$ MeV) + $^{1}$H. The same analysis was performed for all beam/target combinations. If necessary, the calibrations were repeated to cope with time-dependent detector responses, the different energy-loss values in the different target thicknesses and different angular distributions of the fragments.

A different method was used to determine the detection efficiency of the TWIN detector in the experiment with the $^{238}$U beam. This method proved to be more precise in this case than the one described above due to the larger fissility of this nucleus. As mentioned before, the coincidence spectrum recorded with the TWIN detector, Figure 4, does not allow recognising all fission events. The missing events are characterized by trajectories close to the horizontal plane in which two fragments are passing through the same part of the TWIN chamber. In contrast, the second ionization chamber MUSIC2 placed behind the target registers all reaction products without any losses, although it does not measure the multiplicity of the event. Most fission products, which are not recognized by the TWIN chamber, are emitted with sizeable horizontal angles with respect to the beam axis. As the four anodes of the MUSIC2 ionisation chamber are positioned on the side of the detector, it was possible to obtain the information on horizontal angle from a multiple measurement of the drift times. Due to the selective sensitivity of the electronics to the first signal arriving, only the fission fragment with the trajectory closest to the anode of the second ionization chamber is registered. In Figure 11, the energy loss recorded in MUSIC2 is shown as a function of the raw drift-time differences between the first and the fourth anode of MUSCI2 for the events outside the polygon window of Figure 4. Figure 11 proves that fission fragments not identified by TWIN detector as such can be distinguished from fragmentation products by their large horizontal angles.

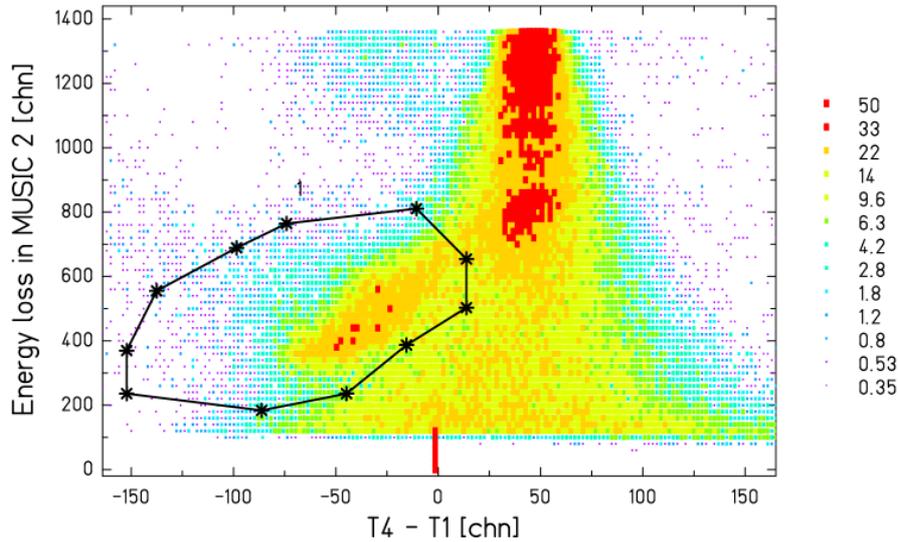

**Figure 11: Energy loss recorded in the ionisation chamber MUSIC2 behind the target as a function of the raw drift-time differences between the first and the forth anode of this detector for the system $^{238}$U, 1 $A$ GeV on $(CH_2)_n$. The spectrum is collected for events outside the window shown in Figure 4. Since the different delays of the two signals (T1 and T4) where not synchronised, the beam direction does not correspond to T4-T1=0 but to T4-T1≈45. The polygon window marks events which are clearly recognized as fission fragments due to their smaller drift-time differences, corresponding to larger emission angles. For energy-loss signals below channel 100, no drift time was recorded.**

Figure 12 demonstrates how this information, in combination with the information from the TWIN detector, was used to improve the detection efficiency of the set-up for fission fragments and, in particular, to determine the detection efficiency of the TWIN detector. The black histogram shows the energy-loss spectrum of all reaction products registered in the second ionisation chamber MUSIC2. The broad peak corresponding to fission fragments between channels 300 and 900 is clearly visible. However, there is also a continuous background of events from other reactions. For determining this background quantitatively, first those events were subtracted, which were recognized as fission by the TWIN detector, Figure 4, resulting in the red histogram. Still, this histogram contains those fission events, which were not identified by the TWIN chamber due to its limited efficiency. In a next step, we subtracted those events, which fell into the condition characterized in Figure 11 as fission events by the ionization chamber behind the target, resulting in the blue histogram. The blue histogram shows a continuous behaviour up to channel 600, proving that essentially all light fission fragments have been identified at this stage. Only some heavy fission fragments were not yet discarded due to their small emission angles. A forth-order polynomial, fitted to the regions below channel 600 and above channel 1000 represents the estimated background of non-fission reaction products. With respect to this estimate on the background, we can deduce that the TWIN chamber, Figure 4, recognizes (89 ± 3) % of the fission events in this case. The additional use of the ionisation chamber directly behind the target, Figure 11, raises the detection efficiency for fission fragments to ($98.7^{+1.3}_{-2}$) %. These values refer to the $^{238}$U beam at about 1 $A$ GeV. For the lower energy of about 500 $A$ MeV, the detection efficiency for fission fragments of the TWIN chamber is (92 ± 3) %.

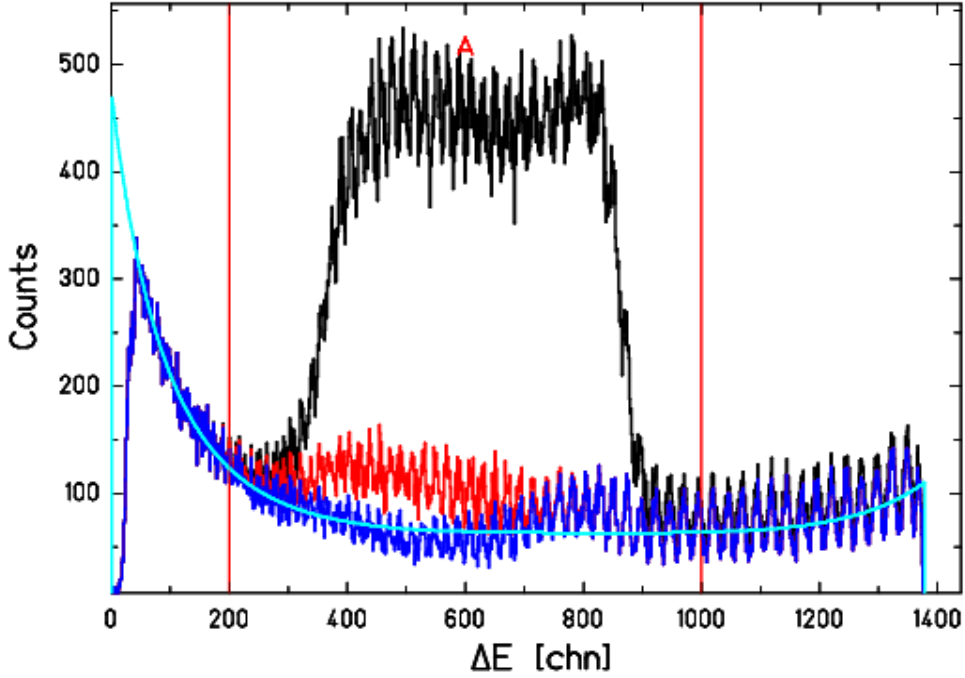

**Figure 12: Energy-loss signal of the ionisation chamber behind the target for the system $^{238}$U (1 $A$ GeV) + (CH$_2$)$_n$. The different histograms denote all reaction products (black histogram) and reaction products, which were not recognized as fission products by the TWIN detector, see Figure 4, (red histogram). The blue histogram was accumulated with the additional condition, that the reaction product was not recognized as fission product by the two-dimensional spectrum shown in Figure 11. The fluctuations in the histograms correspond to the different elements recorded. The light blue curve is a forth-order polynomial, fitted to the last histogram, excluding the region between channels 600 and 1000. It represents the background of all reaction products, except fission.**

## 4. Results

The numerical results of the measurement were evaluated for several periods of the experiments. Measurements with the $^{208}$Pb beam were performed with a hydrogen target, a deuterium target, and an empty target container. For the measurements with the $^{238}$U beam, a plastic target (CH$_2$)$_n$ and a carbon target as well as a measurement without a target inserted were performed. The target thicknesses are given in Table 1. As mentioned above, the number of valid projectiles ($N_p$) properly hitting the target was determined from the condition shown in Figure 2, requiring no response from the veto detector. The number of fission events ($N_f$) was determined from the condition shown in Figure 4. The fluctuations of the data points are in agreement with the expectations due to the statistical uncertainties, given by Poisson statistics according to the observed number of counts. The final results for the measured relative fission yields are listed in Table 2. All statistical uncertainties are below 1%.

The largest contributions to the uncertainties are due to uncertainties of the target thickness, secondary reactions, and the detection efficiency $\varepsilon_{TWIN}$ of the TWIN chamber. The largest correction is due to the attenuation of the beam intensity in the target. After passing half the

target[§§], a fraction of 5.2 % or 6.6 %, of the $^{208}$Pb projectiles undergo nuclear reactions in the hydrogen or the deuterium target, respectively. From calculations with the ABRABLA nuclear-reaction code [19,12,20], we deduce that about half of the reaction products are not fissile any more. This requires a corresponding correction $\varepsilon_{beam}$ of about 2.6 % or 3.3 % to the cross section for the hydrogen or the deuterium target, respectively. The corresponding reaction rate for the $^{238}$U beam is 3.3 % in half the plastic and the carbon target. In this case, most of the reaction products, which consist of fission fragments to a great part, are not fissile any more. Thus, we apply a correction $\varepsilon_{beam}$ of 3 % in the case of the $^{238}$U beam.

Nuclear reactions of the fission fragments in the second part of the target and in consecutive layers before they reach the TWIN chamber are less important, since most of them are peripheral, and thus most of the reaction products are heavy enough to be still identified as fission fragments. In the $^{208}$Pb experiment, reactions in the hydrogen or deuterium target lead only to moderate mass losses, and only relatively central collisions with heavier nuclei in consecutive layers, mostly air, induce some losses. These were estimated to 1/3 of the total nuclear reaction rate of 3 % (see Table 1) for each fragment. This results in a correction $\varepsilon_{frag}$ of 2 %, since the loss of one fragment is sufficient not to recognise the fission event any more. In the $^{238}$U experiment, the probability for a "central" reaction of a fission fragment in the second half of the carbon target is 1 %, the corresponding probability in the second half of the plastic target is 0.3 %. This leads to the slightly larger corrections $\varepsilon_{frag}$ of 4 % for the carbon target and 2.6 % for the plastic target on the whole path to the TWIN chamber. Table 2 collects all values, which are relevant for determining the total fission cross sections of the systems investigated.

**Table 1: Thicknesses and reaction rates for the different targets and the layers of matter located between the target and the double IC. The relative nuclear reaction rates in the hydrogen and deuterium targets were calculated for $^{208}$Pb at 500 $A$ MeV. The relative nuclear reaction rates in the other layers have been calculated for $^{238}$U at 1 $A$ GeV, while the relative nuclear reaction rates in the rest of the layers were calculated for the nucleus Z = 46 A = 116 at 1 $A$ GeV. The energy dependence of the reaction rate is negligible for our purpose. All calculations have been performed using the model of [21].**

| | Material | Thickness [mg/cm$^2$] | Relative nuclear reaction rate |
|---|---|---|---|
| Targets | $^1$H | 87.3 ± 2.2 | 0.103 |
| | $^2$H | 201 ± 5 | 0.133 |
| | (CH$_2$)$_n$ | 198.3 ± 0.2 | 0.067 |
| | C | 376.0 ± 0.9 | 0.066 |
| Layers between target and double IC | Air | 23.9 | 0.003 |
| | Ar (MUSIC) | 99.7 | 0.005 |
| | Air | 197.9 | 0.022 |

The result of the above-detailed analysis in terms of cross sections is presented in Table 3. The relative uncertainties given for the reactions $^{208}$Pb+$^1$H, $^2$H include also a small contribution of 1.5% coming from misidentification of angles in upper TWIN part, as discussed in Section 3.3. In the case of the $^{238}$U experiment, the total fission cross section with hydrogen is deduced from the values obtained for plastic and carbon target as: $\sigma(H) = (\sigma(CH_2)-\sigma(C))/2$. The data are compared with results obtained by summing up the individual nuclide cross sections determined in experiments measured at the GSI fragment separator FRS (column 3) and with the systematics of Prokofiev (column 4) [13]. There is fair agreement to the previous experimental results within the corresponding experimental uncertainties, except for the case of $^{208}$Pb on $^{1}$H at 500 $A$ MeV, where the present value is considerably lower. The new value is in much better agreement with the systematics. Results with uncertainties comparable to the ones of the present experiment have been published recently by Kotov et al. [22]. Two data points for the system $^{238}$U + $^{1}$H, which can directly be compared, agree well with our results within the experimental uncertainties, as displayed in Table 3.

---

[§§]) Assuming linear decrease of the beam intensity along the target thickness, the mean beam attenuation is equal to the value after the passage of half the target.

**Table 2: Measured relative fission yields ($n_f = N_f/N_p$), corrections for TWIN efficiencies, beam attenuation in the target and destruction of fission fragments before reaching the TWIN detector. The total relative fission yield $n_f^{tot}$ represents relative fission yields corrected for TWIN efficiency, beam attenuation and destruction of fragments:** $n_f^{tot} = \dfrac{n_f}{\varepsilon_{TWIN} \cdot (1-\varepsilon_{beam})(1-\varepsilon_{frag})}$, **while net relative fission yield $n_f^{net}$ is obtained after subtracting the contribution from the empty target:** $n_f^{net} = n_f^{tot}(full) - n_f^{tot}(empty)$.

| Reaction | Relative fission yield $n_f$ | TWIN efficiency $\varepsilon_{TWIN}$ [%] | Beam attenuation $\varepsilon_{beam}$ [%] | Destruction of fragments $\varepsilon_{frag}$ [%] | Total relative fission yield $n_f^{tot}$ | Net relative fission yield $n_f^{net}$ |
|---|---|---|---|---|---|---|
| $^{208}$Pb (500 $A$ MeV) + $^{1}$H | 0.00815 | 94 ± 3 | 2.6 ± 2 | 2 ± 1 | 0.00908 | 0.00763 |
| $^{208}$Pb (500 $A$ MeV) + $^{2}$H | 0.01217 | 94 ± 3 | 3.3 ± 2 | 2 ± 1 | 0.01366 | 0.01221 |
| $^{208}$Pb (500 $A$ MeV) + empty | 0.00133 | 94 ± 3 | ≈ 0.5 | 2 ± 1 | 0.00145 | --- |
| $^{238}$U (545 $A$ MeV) + (CH$_2$)$_n$ | 0.0440 | 92 ± 3 | 3 ± 2 | 2.6 ± 1.5 | 0.0506 | 0.0391 |
| $^{238}$U (545 $A$ MeV) + C | 0.0357 | 92 ± 3 | 3 ± 2 | 4 ± 2 | 0.0417 | 0.0302 |
| $^{238}$U (545 $A$ MeV) + empty | 0.0103 | 92 ± 3 | ≈ 0.5 | 2 ± 1 | 0.0115 | --- |
| $^{238}$U (935 $A$ MeV) + (CH$_2$)$_n$ | 0.0404 | 89 ± 3 | 3 ± 2 | 2.6 ± 1.5 | 0.0480 | 0.0381 |
| $^{238}$U (935 $A$ MeV) + C | 0.0299 | 89 ± 3 | 3 ± 2 | 4 ± 2 | 0.0361 | 0.0262 |
| $^{238}$U (935 $A$ MeV) + empty | 0.0086 | 89 ± 3 | ≈ 0.5 | 2 ± 1 | 0.0099 | --- |

**Table 3: Total fission cross sections determined in the present work are compared to previous results. Energy values correspond to those at the target entrance.**

| Reaction | $\sigma_{tot}$ (present work) | $\sigma_{tot}$ (FRS data) | Prokofiev systematics [13] | Kotov et al. [22] |
|---|---|---|---|---|
| $^{208}$Pb (500 $A$ MeV) + $^{1}$H | (146 ± 7) mb | (232 ± 33) mb [23] | 112 mb | |
| $^{208}$Pb (1 $A$ GeV) + $^{1}$H | | (163 ± 26) mb [24] | 116 mb | |
| $^{208}$Pb (500 $A$ MeV) + $^{2}$H | (203 ± 9) mb | | Not included | |
| $^{208}$Pb (1 $A$ GeV) + $^{2}$H | | (169 ± 14) mb [25] | Not included | |
| $^{238}$U (505 $A$ MeV) + $^{1}$H | | | 1.38 b | (1.491 ± 0.078) b |
| $^{238}$U (545 $A$ MeV) + $^{1}$H | (1.49 ± 0.10) b | | 1.36 b | |
| $^{238}$U (935 $A$ MeV) + $^{1}$H | (1.55 ± 0.10) b | | 1.28 b | |
| $^{238}$U (1 $A$ GeV) + $^{1}$H | | (1.53 ± 0.20) b [26] | 1.27 b | (1.489 ± 0.064) b |
| $^{238}$U (1 $A$ GeV) + $^{2}$H | | (2.00 ± 0.42) b [27] | Not included | |

## 5. Discussion and conclusion

In the present work, new data on total fission cross sections in spallation reactions have been obtained in a dedicated experiment in inverse kinematics at the GSI experimental facility. High precision between 5% and 7% could be achieved only due to the fact that the experimental set-up detects the fission fragments with an efficiency close to 100% and that fission events are unambiguously identified and distinguished from other reactions by the dedicated high-resolution detection system. In addition, the projectiles are individually counted using the same detectors. The raw experimental results given by the number of projectiles and fission fragments directly recorded are very close to the final cross-section values. The corrections which have to be applied are thus only a few and amount just to a few percents.

Previously measured data using the high-resolution spectrometer FRS are also listed in table 3 for comparison. Those data are much more detailed, since they provide the production cross sections of all individual nuclides and their velocity distributions. The total fission cross sections are obtained by summing up the individual contributions. The precision on the absolute cross section is not so high due to the large corrections applied for the losses caused by the limited angular acceptance of the spectrometer. The present results for the systems $^{208}$Pb (500 $A$ MeV) + $^{1}$H and $^{238}$U (1 $A$ GeV) + $^{1}$H can be directly compared to the FRS data. While the values for the second system agree very well, there is a severe discrepancy for the first one. The reason for this deviation is not clear in the moment. The systems $^{208}$Pb (1 $A$ GeV) + $^{1}$H and $^{208}$Pb (500 $A$ MeV) + $^{2}$H introduce the same total energy into the system, but the fission cross sections are not the same: The deuteron-induced reactions lead to an appreciably higher fission cross section.

It is interesting to note that the experiment performed by Kotov et al. [22], which agrees well with our result, was done with some similar characteristics as the present experiment, although direct kinematics was used. Essential features, which ensure the high quality of the experiment, were direct counting of the beam particles and large angular acceptance of the set-up for the fission fragments.

The cross-section values obtained in the present work are on the average appreciably larger than the predictions of Prokofiev's systematic, which is based on the evaluation of all experimental data available before 2001. The deviation is between 20 % and 30 % for the spallation-fission reaction of $^{208}$Pb induced by 500 MeV protons and for the spallation-fission reaction of $^{238}$U induced by 1 GeV protons, respectively. The value for the spallation-fission reaction of $^{238}$U induced by 500 MeV protons, however agrees within the experimental uncertainty. Thus, there is no general trend; the present data reveal deficiencies of the systematics in both the dependencies on the mass of the system and on the beam energy in the order of 20 % to 30 %.

We conclude that additional experimental effort is needed to improve the general knowledge of the total fission cross sections. It seems that the new approach introduced in the present work can provide an essential contribution to this goal. The agreement with a recent experiment in direct kinematics has proven that also such experiments can reach high precision, when uncertainties on the beam dose and on the detection efficiency of the fission detectors are well controlled.

## Acknowledgements

This work has been supported by the European Commission under contract no. 012985 (EURATOM Intra-European Fellowship FISA2004) and contract no. 516520-FI6W (EURATOM Integrated Project EUTRANS). The EC is not liable for any use that may be made of the information contained herein.

## References

[1] C. D. Bowman, Ann. Rev. Nucl. Part. Sci. 48 (1998) 505.
[2] C. Rubbia et al., Report CERN/AT/95-44/(ET), 1995.

[3] H. Nifenecker, S. David, J. M. Loiseaux, O. Meplan, Nucl. Instrum. Methods A 463 (2001) 428.
[4] R. Michel, P. Nagel, OECD/NEA, Paris, (1997) NSC/DOC(97)-1, NEA/P&T No14.
[5] http://www.gsi.de/tramu/
[6] http://nds121.iaea.org/alberto/mediawiki-1.6.10/index.php/Main_Page
[7] http://www.fynu.ucl.ac.be/collaboration/hindas/
[8] http://pceet075.cern.ch/
[9] http://nuklear-server.fzk.de/eurotrans/Start.html
[10] http://www.gsi.de/charms
[11] A. Boudard, J. Cugnon, S. Leray, C. Volant, Phys. Rev. C 66 (2002) 044615.
[12] A. Kelić, M.V. Ricciardi, K.-H. Schmidt, Proceedings of the Joint ICTP-IAEA Advanced Workshop on Model Codes for Spallation Reactions, 4-8 February 2008, ICTP Trieste, Italy, IAEA INDC(NDS)-530 (2008) 181-221.
[13] A. V. Prokofiev, Nucl. Instrum. Methods A 463 (2001) 557.
[14] A. V. Prokofiev, "Nucleon-induced fission cross sections of heavy nuclei in the intermediate energy region", PhD Thesis, Uppsala University, 2001.
[15] H. Geissel, P. Armbruster, K.-H. Behr, A. Bruenle, K. Burkard, M. Chen, H. Folger, B. Franczak, H. Keller, O. Klepper, B. Langenbeck, F. Nickel, E. Pfeng, M. Pfützner, E. Roeckl, K. Rykaczewsky, I. Schall, D. Schardt, C. Scheidenberger, K.-H. Schmidt, A. Schroeter, T. Schwab, K. Suemmerer, M. Weber, G. Muenzenberg, T. Brohm, H.-G. Clerc, M. Fauerbach, J.-J. Gaimard, A. Grewe, E. Hanelt, B. Knoedler, M. Steiner, B. Voss, J. Weckenmann, C. Ziegler, A. Magel, H. Wollnik, J.-P. Dufour, Y. Fujita, D. J. Vieira, B. Sherrill, Nucl. Instrum. Methods B 70 (1992) 286.
[16] B. Voss, T. Brohm, H.-G. Clerc, A. Grewe, E. Hanelt, A. Heinz, M. de Jong, A. Junghans, W. Morawek, C. Roehl, S. Steinhaeuser, C. Ziegler, K.-H. Schmidt, K.-H. Behr, H. Geissel, G. Münzenberg, F. Nickel, C. Scheidenberger, K. Sümmerer, A. Magel, M. Pfützner, Nucl. Instrum. Methods A 364 (1995) 150.
[17] M. Pfützner, H. Geissel, G. Münzenberg, F. Nickel, C. Scheidenberger, K.-H. Schmidt, K. Sümmerer, T. Brohm, B. Voss, H. Bichsel, Nucl. Instr. Meth. B 86 (1994) 213.
[18] K.-H. Schmidt, S. Steinhäuser, C. Böckstiegel, A. Grewe, A. Heinz, A. R. Junghans, J. Benlliure, H.-G. Clerc, M. de Jong, J. Müller, M. Pfützner, B. Voss, Nucl. Phys. A 665 (2000) 221
[19] M. V. Ricciardi, P. Armbruster, J. Benlliure, M. Bernas, A. Boudard, S. Czajkowski, T. Enqvist, A. Kelić, S. Leray, R. Legrain, B. Mustapha, J. Pereira, F. Rejmund, K. -H. Schmidt, C. Stéphan, L. Tassan-Got, C. Volant, O. Yordanov, Phys. Rev, C 73 (2006) 014607.
[20] A. R. Junghans, M. de Jong, H.-G. Clerc, A.V. Ignatyuk, G.A. Kudyaev, K.-H. Schmidt, Nucl. Phys. A 629 (1998) 635.
[21] C. J. Benesh, B. C. Cook, J. P. Vary, Phys. Rev. C 40 (1989) 1198.
[22] A. A. Kotov, L. A. Vaishnene, V. G. Vovchenko, Yu. A. Gavrikov, V. V. Poliakov, M. G. Tverskoy, O. Ya. Fedorov, Yu. A. Chestnov, A. I. Shchetkovskiy, A. V. Shvedchikov, A. Yu. Doroshenko, and T. Fukahori, Phys. Rev. C 74 (2006) 034605.
[23] B. Fernandez, P. Armbruster, L. Audouin, J. Benlliure, M. Bernas, A. Boudard, E. Casarejos, S. Czajkowski, J. E. Ducret, T. Enqvist, B. Jurado, R. Legrain, S. Leray, B. Mustapha, J. Pereira, M. Pravikoff, F. Rejmund, M. V. Ricciardi, K.-H. Schmidt, C. Stéphan, J. Taïeb, L. Tassan-Got, C. Volant, W. Wlazlo, Nucl. Phys. A 747 (2005) 227.
[24] T. Enqvist, W. Wlazlo, P. Armbruster, J. Benlliure, M. Bernas, A. Boudard, S. Czajkowski, R. Legrain, S. Leray, B. Mustapha, M. Pravikoff, F. Rejmund, K.-H. Schmidt, C, Stéphan, J. Taïeb, L. Tassan-Got, C. Volant, Nucl. Phys. A 686 (2001) 481.

[25] T. Enqvist , P. Armbruster, J. Benlliure, M. Bernas, A. Boudard, S. Czajkowski, R. Legrain, S. Leray, B. Mustapha, M. Pravikoff, F. Rejmund, K.-H. Schmidt, C. Stéphan, J. Taïeb, L. Tassan-Got, F. Vivès, C. Volant, W. Wlazlo, Nucl. Phys. A 703 (2002) 435.

[26] M. Bernas, P. Armbruster, J. Benlliure, A. Boudard, E. Casarejos, S. Czajkowski, T. Enqvist, R. Legrain, S. Leray, B. Mustapha, P. Napolitani, J. Pereira, F. Rejmund, M. V. Ricciardi, K.-H. Schmidt, C. Stéphan, J. Taïeb, L. Tassan-Got, C. Volant, Nucl. Phys. A 725 (2003) 213.

[27] J. Pereira, J. Benlliure, E. Casarejos, P. Armbruster, M. Bernas, A. Boudard, S. Czajkowski, T. Enqvist, R. Legrain, S. Leray, B. Mustapha, M. Pravikoff, F. Rejmund, K.-H. Schmidt, C. Stéphan, J. Taïeb, L. Tassan-Got, C. Volant, W. Wlazlo, Phys. Rev. C 75 (2007) 014602.